# SKEIN SPACES AND SPIN STRUCTURES

## John W. Barrett

### 10 March 1997

Abstract. This paper relates skein spaces based on the Kauffman bracket and spin structures. A spin structure on an oriented 3-manifold provides an isomorphism between the skein space for parameter $A$ and the skein space for parameter $-A$.

There is an application to Penrose's binor calculus, which is related to the tensor calculus of representations of SU(2). The perspective developed here is that this tensor calculus is actually a calculus of spinors on the plane, and the matrices are determined by a type of spinor transport which generalises to links in any 3-manifold.

A second application shows that there is a skein space which is the algebra of functions on the set of spin structures for the 3-manifold.

This paper relates skein spaces based on the Kauffman bracket and spin structures. The main result for a general parameter $A$ is that a spin structure on an oriented 3-manifold provides an isomorphism between the skein spaces for parameters $\pm A$.

Specialising to the case of $A = \pm 1$ gives the application to Penrose's binor calculus. The binor calculus is related to a tensor calculus of invariants for the group SU(2). The perspective developed here is that this tensor calculus is actually a calculus of spinors on the plane, and the matrices are determined by a type of spinor transport which generalises to links in any 3-manifold.

As an elementary example, the unknot corresponds to the operation of transporting a spinor in $\mathbb{C}^2$ one full turn around a circle. This operation on spin space is minus the identity in SU(2), which has trace $-2$, the Kauffman bracket evaluation for the unknot for parameter $A = \pm 1$.

However, the binor calculus is related to $A = -1$, whereas the geometrical description in terms of spinors occurs for $A = 1$. The isomorphism of the two skein spaces provides the relation between these.

More generally, the skein spaces for $A^6 = 1$ have a quotient which is a commutative algebra. For $A = -1$ this is known to be related to the algebra of functions on the space of flat SU(2)-connections on $M$. The analogous description for $A = 1$ is given here in terms of the flat connections over the frame bundle, where the holonomy around the fibres of the frame bundle is non-trivial.

In the case of a primitive cube root of 1, the commutative skein algebra is the algebra of functions on the set of spin structures of the manifold $M$. This result follows naturally by using the isomorphism with the skein space for $-A$, a primitive sixth root of 1, for which it is easily shown that the algebra is the algebra of functions on $H^1(M, Z_2)$.

**Skein space.**

A framed link in an oriented differentiable 3-manifold is a disjoint set of embedded circles, each having a non-zero normal vector field.

Let $M$ be an oriented 3-manifold. For a parameter $A \in \mathbb{C}$, $A \neq 0$, the skein space $\mathrm{Skein}_A(M)$ is the vector space over $\mathbb{C}$ generated by the framed links, subject to the relations

(1) Ambient isotopy of the link.

The following two relations apply to framed links which differ in a region which is isomorphic to $D^2 \times [0,1]$ by an orientation-preserving map. The diagrams show the projection onto $D^2$. The normal vector field for each curve is along the fibres of the projection, in the direction in which the coordinate increases.

(2) For a component of the link which is an unlinked unknot, this can be removed, scaling the remaining link by an element of $\mathbb{C}$.

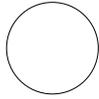
$$= -A^2 - A^{-2}$$

(3) The Kauffman bracket relation.

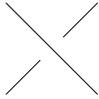
$$= A$$

A similar definition is given in [Hoste and Przytycki 1992]. As an example of this definition, if $M = \Sigma \times [0,1]$, then the skein space can be regarded as the space obtained by applying the relations directly to the link diagrams on the orientable surface $\Sigma$. The Kauffman bracket for link diagrams on the plane is given in [Kauffman 1987]. These constructions are used in topological quantum field theory in [Roberts 1994].

Now let $M$ be a manifold with a spin structure $s$. Several different characterisations of a spin structure are given by [Milnor 1963] and [Kirby 1989]. Let $\gamma$ be a framed embedded circle. There is a uniquely determined homotopy class of trivialisations of $TM$ restricted to $\gamma$. Define $\mathrm{Spin}(\gamma, s) \in \mathbb{Z}_2$ to be 1 if the homotopy class agrees with that determined by the spin structure, otherwise 0. The homotopy class can be described explicitly, given a choice of an orientation for the circle. The first vector field is the normal vector field, the second vector field is the tangent vector field, and the third vector field is determined by the orientation of $M$ to give an oriented frame. Changing the orientation of the circle changes the second and third vector fields, but this trivialisation is homotopic to the original one by a rotation of axes. In this way, $\gamma$ inherits a spin structure from $M$. Then $\mathrm{Spin}(\gamma, s)$ is 0 if this spin structure on the circle bounds a spin structure on a disk, otherwise it is 1.

The spin structure on $M$ can be used to 'flip' $A$ to $-A$.

**Theorem 1.** *Each spin structure $s$ for $M$ determines a linear map*

$$\phi(s) \colon \mathrm{Skein}_A(M) \to \mathrm{Skein}_{(-A)}(M)$$

*by multiplying each link l by*

$$(-1)^{\sum \text{Spin}(\gamma, s)}$$

*where the sum is over the components $\gamma$ of the link $l$.*

*Proof.* Relations (1) and (2) are unaffected by the change $A \to -A$. The map $\phi(s)$ is the same on both sides of these relations, the contractible knot in (2) having $\text{Spin}(\gamma, s) = 0$.

For relation (3), fix a Riemannian metric on $M$ and consider the bundle $E$ of orthonormal oriented frames. The normal vector field can be made orthogonal to the tangent vector and normalised to unit length. The spin structure $s$ is a cohomology class $s \in H^1(E, \mathbb{Z}_2)$. Choosing an orientation for a framed embedded circle $\gamma$ in $M$ determines a unique lift $\tilde{\gamma}$ in $E$. Then $\text{Spin}(\gamma, s) = s(\tilde{\gamma}) + 1 \in \mathbb{Z}_2$. The difference of the two values of $\sum \text{Spin}(\gamma, s)$ for two of the terms in the skein relation can be found by computing the value of $s$ on the difference of the corresponding homology classes in $E$.

There are three cases, depending on the way that the four boundary points in the skein relation are connected to each other by the link. Only two of these cases are essentially different.

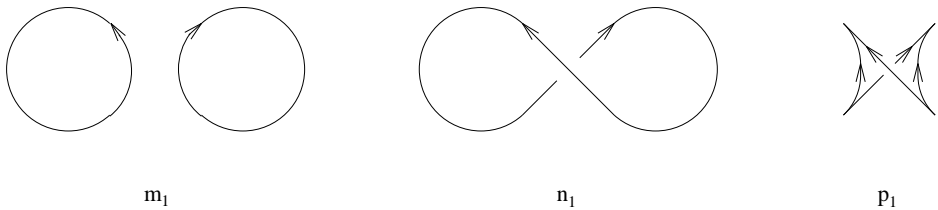

$$m_1 \qquad\qquad n_1 \qquad\qquad p_1$$

Figure 1

*Case A.* Figure 1 shows two terms in the skein relation, $m_1$ and $n_1$, which have been closed into loops on the diagram to indicate which ends are joined by the sublinks. Orientations for these have been chosen so that they coincide outside of the region isomorphic to $D^2 \times [0, 1]$. Accordingly, the lifts $\tilde{m}_1$ and $\tilde{n}_1$ to the frame bundle also coincide there. Therefore the difference in their homology classes is just the homology class of the lift of $p_1$, which consists of oriented differentiable segments. The lift of $p_1$ is a continuous curve and its homology class is zero. Therefore $s(\tilde{m}_1) = s(\tilde{n}_1)$, and so the values of $\sum \text{Spin}(\gamma, s)$ for the two links are different elements of $\mathbb{Z}_2$. This is because the number of components in each link differs by one.

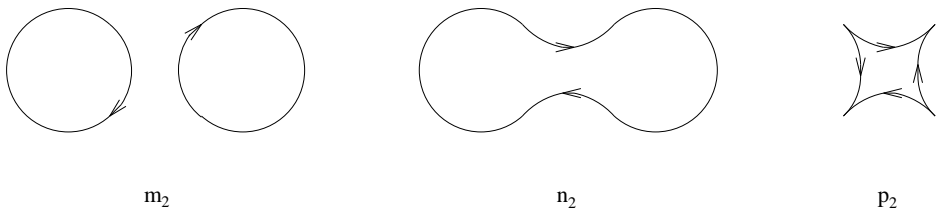

$$m_2 \qquad\qquad n_2 \qquad\qquad p_2$$

Figure 2

Figure 2 shows another pair of terms in the skein relation. For this pair, $s(\tilde{m}_2) = s(\tilde{n}_2) + s(\tilde{p}_2)$, but this time the homology class of $\tilde{p}_2$ is the generator of $H_1(\mathrm{SO}(3), \mathbb{Z}_2)$. Therefore $s(\tilde{p}_2) = 1$ and so the values of $\sum \mathrm{Spin}(\gamma, s)$ for each link are equal.

The calculation with Figures 1 and 2 shows that the sign of the map $\phi(s)$ is the same for the two terms on the right hand side of the skein relation but is different for the term on the left. Therefore the relation for parameter $A$ is mapped to the relation for parameter $-A$.

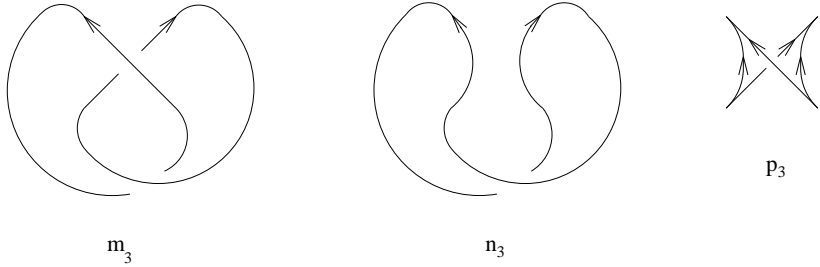

$m_3$              $n_3$              $p_3$

FIGURE 3

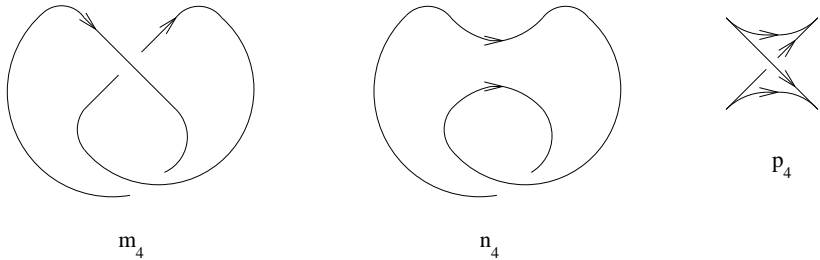

$m_4$              $n_4$              $p_4$

FIGURE 4

*Case B.* Figures 3 and 4 show two pairs of terms in the skein relation. The argument is the same as for case A, except that now both $p_3$ and $p_4$ have lifts with homology zero. Therefore in both pairs the sign of the map $\phi(s)$ is different for the two terms, and again it follows that $\phi(s)$ has the opposite sign on the term on the left of the skein relation to the two terms on the right.

In a similar way, each cohomology class $h \in H^1(M, \mathbb{Z}_2)$ determines an endomorphism $\phi(h)$ of $\mathrm{Skein}_A(M)$ by multiplying each link by

$$(-1)^{\sum h(\gamma)}.$$

If $s_1$ and $s_2$ are two spin structures for $M$, then $\phi(s_2) = \phi(h) \circ \phi(s_1)$, where $h$ is the unique element which is mapped to $s_2 - s_1$ by $\pi^* \colon H^1(M, \mathbb{Z}_2) \to H^1(E, \mathbb{Z}_2)$.

**Binors and spinors.**

For the values $A = \pm 1$, the over- and under-crossings coincide. For the value $A = -1$, the relations coincide with those of Penrose's binor calculus [Penrose 1979], as explained by [Kauffman 1990]. For diagrams on the plane, the skein space is one-dimensional, and can be identified with $\mathbb{C}$ by identifying the empty diagram with 1. This is equivalent to using the formula

$$(-2)^{\#\text{circles}}.$$

This formula generalises to any $M$, giving a linear functional $P\colon \text{Skein}_{(-1)}(M) \to \mathbb{C}$.

The linear functional $P$ can be twisted by $h \in H^1(M; \mathbb{Z}_2)$ to give $P^h\colon \text{Skein}_{(-1)} \to \mathbb{C}$ or by a spin structure $s$ to give $P_s\colon \text{Skein}_1 \to \mathbb{C}$. The formula for $P_s$ is

$$l \mapsto (-2)^{\#\text{circles}}(-1)^{\sum \text{Spin}(\gamma, s)}$$

for a link $l$.

For $M = \mathbb{R}^2 \times [0, 1]$, this formula becomes

$$(-2)^{\#\text{circles}}(-1)^{\#\text{crossings}}$$

for the unique spin structure $s$. In this formula, #crossings denotes the number of crossing points in the corresponding planar diagram. This follows because $\text{Spin}(\gamma, s)$ is the number of self-crossings modulo 2, and the number of crossings of distinct components is even.

Moussouris [1979] explains the relation between the Penrose evaluation $P$ for planar diagrams and the evaluation of tensor invariants for SU(2), which was Penrose's motivation for studying the binor calculus. From the point of view explored here, it is actually $P_s$ which has the natural description in terms of tensor calculus.

Moussouris' computation for $P$ can be adjusted, in the light of the preceding paragraphs, to give a description of $P_s$. It amounts to assigning tensors to planar diagrams with free ends, generated by

 $\rightarrow \epsilon_{ab}$

 $\rightarrow \delta_a^b$

where $a, b, c, d = 0, 1$ label basis vectors for $\mathbb{C}^2$ for superscripts and its dual space for subscripts. The tensor components are given by the matrices

$$\epsilon_{ab} = \begin{pmatrix} 0 & 1 \\ -1 & 0 \end{pmatrix}, \qquad \epsilon^{ab} = \begin{pmatrix} 0 & -1 \\ 1 & 0 \end{pmatrix}, \qquad \delta_a^b = \begin{pmatrix} 1 & 0 \\ 0 & 1 \end{pmatrix}$$

with $a$ labelling the column and $b$ labelling the row in each case.

Moussouris starts with the definition of $\epsilon^{ab}$ as the negative of that given here but later includes an extra factor of $(-1)$ into the evaluation. He also gives the evaluation of the crossing as $-\delta_a^d \delta_b^c$, the difference in sign to that given here being accounted for by the difference between $P_s$ and $P$. These tensors verify the relations for the skein space for $A = 1$ directly.

This evaluation for $P_s$ can be generalised to any 3-manifold with any spin structure $s$. The spin structure determines a bundle of spinors. The fibre at point $p \in M$ is called the spin space, $S(p)$, and is determined by the representation of $SU(2)$ in $\mathbb{C}^2$. The principal $SU(2)$ bundle can be identified with the bundle $F$ of spin frames.

Above a point $p$ on a framed curve, there are two preferred frames for the tangent space, as described above, one for each tangent direction. Therefore there are four preferred spin frames. These are related by the action of a $\mathbb{Z}_4$ subgroup of $SU(2)$. These four transformations cover the identity and the rotation of a half turn about the first coordinate axis in $SO(3)$. If $\sigma$ is a segment of a framed link, i.e., a framed curve with two ends $x$ and $y$, then there are exactly four preferred liftings of $\sigma$ to a curve in $F$ which are continuous. Let $f_0 \colon \mathbb{R}^2 \to S(x)$, $f_1 \colon \mathbb{R}^2 \to S(y)$ be the frames at each end of a continuous lifting. The formula $f_1 \circ f_0^{-1} \colon S(x) \to S(y)$ determines a linear isomorphism of the spin spaces at the endpoints. This isomorphism gives the generalisation of Moussouris' representation to an arbitrary spin manifold. The following lemma gives its properties.

**Lemma.** *The linear isomorphism $S(x) \to S(y)$ is independent of the choice of the lifting of $\sigma$ to $F$. Composing framed curves gives the composite isomorphism. Two segments both with endpoints $x, y$ which form a framed circle $\gamma$ give an isomorphism $S(x) \to S(x)$. The trace of this isomorphism is*

$$-2(-1)^{\mathrm{Spin}(\gamma, s)}$$

*Proof.* Changing the lifting amounts to replacing $f_0$ and $f_1$ by $f_0 \circ J$ and $f_0 \circ J$, where $J \colon \mathbb{R}^2 \to \mathbb{R}^2$ is in the $\mathbb{Z}_4$ subgroup, and so $f_1 \circ f_0^{-1}$ is unchanged. The second part follows from $(f_2 \circ f_1^{-1}) \circ (f_1 \circ f_0^{-1}) = f_2 \circ f_0^{-1}$. For the last part, if $\mathrm{Spin}(\gamma) = 0$, then $s(\tilde{\gamma}) = 1$ and the linear isomorphism is $-1$ times the identity, with trace $-2$. If $\mathrm{Spin}(\gamma) = 1$, then $s(\tilde{\gamma}) = 0$ and the linear isomorphism is the identity, with trace equal to 2.

It follows from the lemma that taking the product of the trace for each circle gives exactly $P_s$. To compare the linear isomorphisms with Moussouris' matrices, it is necessary to pick a basis for the spin space for every free end of a diagram. The simplest choice is to take one of the four preferred spin frames at each free end. Then, since the linear isomorphism $S(x) \to S(y)$ takes one preferred spin frame to another one, the matrices are in the $\mathbb{Z}_4$ subgroup, and can be taken as

$$\pm \begin{pmatrix} 0 & 1 \\ -1 & 0 \end{pmatrix}, \qquad \pm \begin{pmatrix} 1 & 0 \\ 0 & 1 \end{pmatrix}.$$

These matrices are orthogonal, and there is an invariant inner product which can be used to identify $S(x)$ with its dual. More abstractly, the existence of an invariant inner product, and a real structure, can be traced to the fact that the framing vector is an invariant tangent vector. This gives an element of $S(x) \otimes S(x)$, for each $x$ on the framed curve, which is preserved under the linear isomorphism.

To recover Moussouris' representation for diagrams on $\mathbb{R}^2$, it is sufficient to choose the Euclidean metric and a constant spinor frame. This should be one of the two frames which covers a standard orthonormal frame. The standard orthonormal frame has its first vector normal to the plane, second vector tangent to the plane in a vertical direction on the page, and third vector also tangent, and horizontal. The free ends of a diagram must be everywhere vertical. More generally, the free ends need not be vertical. Then the tensor associated to a segment has components

$$\begin{pmatrix} \cos\theta/2 & \sin\theta/2 \\ -\sin\theta/2 & \cos\theta/2 \end{pmatrix},$$

where $\theta$ is the angle through which the tangent vector rotates from one end of the curve to the other, in an appropriate sense.

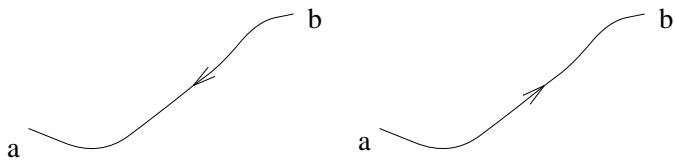

FIGURE 5

The tensor components do not depend on a choice of orientation for the curve. If the components for the left-hand side of Figure 5 are $R_{ab}$ then the components for the right hand side are $R_{ba}^{-1}$. However these are the same as $R$ is an orthogonal matrix.

**Commutative skein algebras.**

A commutative algebra $\mathrm{Skein}'_A(M)$, is defined to be the quotient of $\mathrm{Skein}_A(M)$ by adding the extra relation

(4)

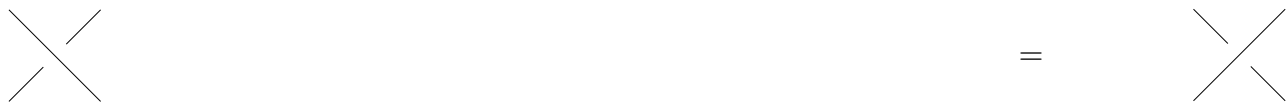

This vector space is an algebra with the product given by the union of links. To define the union, it may be necessary to displace one link so that it does not intersect the other link.

The relation (4) can be expanded with (3) to show that either $A = A^{-1}$, or

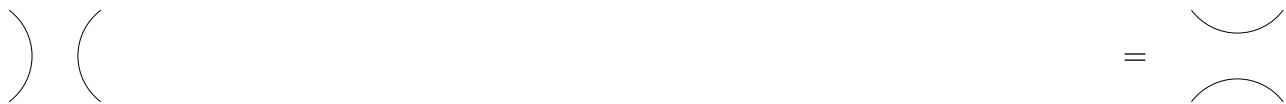

The later relation implies

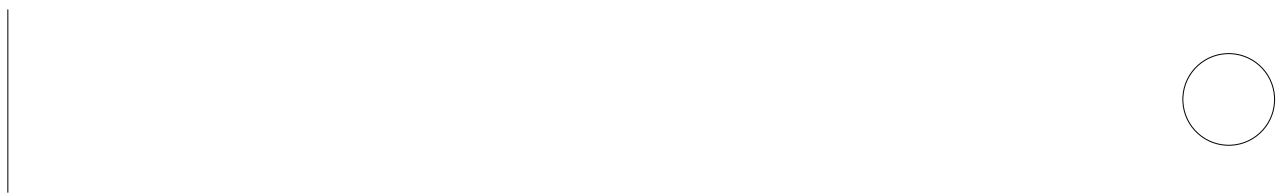

so that every non-empty skein is equivalent to zero unless $-A^2 - A^{-2} = 1$, i.e., $A^2$ is a primitive cube root of 1.

These possibilities for a non-trivial quotient $\mathrm{Skein}'_A(M)$ are exactly the $A$ such that $A^6 = 1$.

**The cases** $A^2 = 1$**.** The extra relation (4) follows from (3), so $\mathrm{Skein}'_A(M) = \mathrm{Skein}_A(M)$. The geometrical interpretation for $A = -1$ is that skein elements provide coordinates for the variety $\mathrm{Ch}(M)$ of characters of $\pi_1(M)$ represented in $\mathrm{SL}(2, \mathbb{C})$. A character is the function $\pi_1 \to \mathbb{C}$ which gives the trace of the representing matrix.

A map $\gamma \colon S^1 \to M$ determines a function on $\mathrm{Ch}(M)$ in the following way. It determines two conjugacy classes in $\pi_1(M)$, $[p]$ and $[p^{-1}]$, on which a character $\chi$ gives a complex number, $-\chi(p) = -\chi(p^{-1})$. This representation of the skein space for $A = -1$ is studied in [Bullock 1996], where it is shown that the coordinate ring of $\mathrm{Ch}(M)$ is a certain quotient of $\mathrm{Skein}_{-1}(M)$.

A character on $M$ is determined by a flat connection on an $\mathrm{SL}(2, \mathbb{C})$ bundle over $M$. For example, the trivial connection determines the Penrose evaluation $P$.

The analogous geometrical description for $A = 1$ is given by considering the characters for the frame bundle $E$. Since $\pi_1$ for the fibre is $\mathbb{Z}_2$, a representation of $\pi_1(E)$ maps the generator for $\pi_1$ of the fibre to one of $\pm 1$ times the identity matrix. Thus $\mathrm{Ch}(E)$ decomposes as $\mathrm{Ch}_1(E) \cup \mathrm{Ch}_{-1}(E)$. The subvariety $\mathrm{Ch}_1(E)$ is naturally isomorphic to $\mathrm{Ch}(M)$ by projection. Therefore elements of $\mathrm{Skein}_{-1}(M)$ provide coordinates for $\mathrm{Ch}_1(E)$.

This allows elements of $\mathrm{Skein}_1(M)$ to be understood as coordinates for $\mathrm{Ch}_{-1}(E)$.

**Theorem 2.** *There is a surjection of* $\mathrm{Skein}_1(M)$ *to the algebra of functions on* $Ch_{-1}(E)$*. A framed embedded circle determines two conjugacy classes in* $\pi_1(E)$*,* $[p]$ *and* $[p^{-1}]$*, and this acts as a function on* $Ch_{-1}(E)$ *by* $\chi \to \chi(p)$*.*

Note the change in sign compared with $\mathrm{Skein}_{-1}(M)$.

*Proof.* Pick a spin structure $s$. This determines a bijection $\mathrm{Ch}_{-1}(E) \to \mathrm{Ch}_1(E)$, given by

$$\chi'(p) = (-1)^{s(p)} \chi(p).$$

Then combining this with $\phi(s)$ of Theorem 1 reduces this to the previously mentioned result of Bullock [1996] on $\mathrm{Skein}_{-1}(M)$.

Let $\gamma$ be a framed embedded circle, which determines $p \in \pi_1(E)$ up to conjugacy and inverse. Then $\phi(s)\gamma = (-1)^{\mathrm{Spin}(\gamma, s)}\gamma \in \mathrm{Skein}_{-1}(M)$, which determines the function $\chi' \mapsto -(-1)^{\mathrm{Spin}(\gamma, s)}\chi'(\gamma)$ on $\mathrm{Ch}_1(E)$, which determines the function

$$\chi \mapsto -(-1)^{s(p)}(-1)^{\mathrm{Spin}(\gamma, s)}\chi(p) = \chi(p)$$

on $\mathrm{Ch}_{-1}(E)$.

**The cases** $1 + A^2 + A^{-2} = 0$**.** The space $\mathrm{Skein}'_A(M)$ is generally a non-trivial quotient of $\mathrm{Skein}_A(M)$, as can be seen for the solid torus.

**Theorem 3.** *For $A$ a primitive sixth root of 1,* $\mathrm{Skein}'_A(M)$ *is isomorphic to the algebra of complex-valued functions on* $H^1(M, Z_2)$*. For $A$ a primitive cube root of 1,* $\mathrm{Skein}'_A(M)$ *is isomorphic to the algebra of complex-valued functions on the set of spin structures of $M$.*

*Proof.* Firstly, when $A$ is a primitive sixth root of 1, there is an algebra isomorphism $\mathrm{Skein}'_A(M) \to \mathbb{C}[H_1(M, Z_2)]$, the group algebra, defined by taking each link to its homology class. This map is well-defined because for these values of $A$, the relations in $\mathrm{Skein}'_A(M)$ do not change the homology class. Conversely, a homology cycle can be represented as an embedded graph with vertices of even order. This is equivalent to a link in a number of different ways by resolving each vertex into pair-wise connections of edges. The skein relations are sufficient to show these are all equal in $\mathrm{Skein}'_A(M)$. The algebra $\mathbb{C}[H_1(M, Z_2)]$ is naturally isomorphic to the algebra of functions on the dual vector space, which is $H^1(M, Z_2)$, by a Fourier transform.

Now $-A$ is a primitive cube root of 1, so Theorem 1 gives an algebra isomorphism of $\mathrm{Skein}'_{-A}(M)$ with the algebra of functions on $H^1(M, Z_2)$, for a choice of spin structure $s$. The set of spin structures on $M$ is naturally identified with $H^1(M, Z_2)$ given the choice $s$. More directly, a link determines a function on the set of spin structures by

$$s \mapsto (-1)^{\sum_\gamma \mathrm{Spin}(\gamma, s)}.$$

**Acknowledgement.** Thanks are due to discussion with Bruce Westbury.

DEPARTMENT OF MATHEMATICS, UNIVERSITY OF NOTTINGHAM, UNIVERSITY PARK, NOTTINGHAM, NG7 2RD, UK
*E-mail address*: jwb@maths.nott.ac.uk